\documentclass{article} 

\usepackage{algorithm}
\usepackage{algorithmic}

\usepackage{graphicx}

\usepackage{amsmath,amsthm,amssymb,enumerate}
\theoremstyle{plain}  
\theoremstyle{definition} 
\theoremstyle{plain}  
\theoremstyle{plain} 

\usepackage{color}

\begin{document}

\title{The cerebellum could solve the motor error problem through error increase prediction}

\author{Sergio Verduzco-Flores \\
Computational Cognitive Neuroscience Laboratory, \\
Department of Psychology and Neuroscience, \\
University of Colorado Boulder, Boulder CO, USA \\
sergio.verduzco@gmail.com
\and
Randall C. O'Reilly \\
Computational Cognitive Neuroscience Laboratory, \\
Department of Psychology and Neuroscience, \\
University of Colorado Boulder, Boulder CO, USA 
}

\date{ }
\maketitle

\begin{abstract}
We present a cerebellar architecture with two main characteristics. The first one
is that complex spikes respond to increases in sensory errors. The second one is that
cerebellar modules associate particular contexts where errors have increased in the
past with corrective commands that stop the increase in error. We analyze our 
architecture formally and computationally for the case of reaching in a 3D environment. 
In the case of motor control, we show that there are synergies of this architecture with
the Equilibrium-Point hypothesis, leading to novel ways to solve the motor error problem.
In particular, the presence of desired equilibrium lengths for muscles provides a way to 
know when the error is increasing, and which corrections to apply.
In the context of Threshold Control Theory and Perceptual Control Theory we show how to 
extend our model so it implements anticipative corrections in cascade control systems 
that span from muscle contractions to cognitive operations.

\end{abstract}

\section{Introduction}

The anatomy of the cerebellum presents a set of well established and striking facts
\cite{eccles_cerebellum_1967,ito_cerebellar_2006}, which have inspired a variety of 
functional theories over the years.  
The cerebellum receives two main input sources, the mossy fibers and
the climbing fibers. The mossy fibers convey a vast amount of afferent and efferent
information, and synapse onto granule cells, Golgi cells, and neurons of the deep
cerebellar nuclei. Granule cells exist in very large numbers, and could be considered
the input layer of the cerebellum; they send axons that bifurcate in the cerebellar cortex,
called parallel fibers, innervating Purkinje cells and molecular layer interneurons.
Purkinje cells have intricate dendritic arbors with about 150 000 parallel fiber connections.
On the other hand, each Purkinje cell receives a single climbing fiber that can provide
thousands of synapses. Activation of a climbing fiber reliably causes a sequence of tightly
coupled calcium spikes, known as a complex spike. In contrast, simple spikes are the action
potentials tonically produced by Purkinje cells, modulated by parallel fiber inputs and feedforward
inhibition from molecular layer interneurons. The sole output from the cerebellar cortex is
constituted by the Purkinje cell axons, which send inhibitory projections to the deep cerebellar nuclei
and to the vestibulum. Cells in the deep cerebellar nuclei can send projections to diverse targets,
such as the brainstem, the thalamus, the spinal cord, and the inferior olivary nucleus. The inferior
olivary nucleus is the origin of climbing fibers, which are the axons of electrotonically-coupled 
olivary cells that experience subthreshold oscillations in their membrane potential.

There is a prevailing view that the cerebellum is organized into modular circuits that 
perform similar computations. Sagittal regions of Purkinje cells called microzones receive climbing
fibers from a cluster of coupled olivary neurons, and tend to be activated by the same functional
stimuli. Purkinje cells in a microzone project to the same group of cells in the cerebellar nuclei,
which in turn send inhibitory projections to the olivary neurons that innervate the microzone.
A microzone together with its associated cerebellar nuclear cells is called a microcomplex, which
together with its associated olivary cells constitutes an olivo-cerebellar module.

In one of the first and most influential theories about cerebellar function, 
developed by a succession of researchers \cite{Marr69,Albus71,ito_climbing_1982}, the 
convergence of mossy fibers (which carry sensory and motor signals into the cerebellum) onto Purkinje cells 
supports pattern recognition in a manner similar to a perceptron. This pattern recognition
capacity is used to improve motor control, and the Marr-Albus-Ito hypothesis 
states that the other major cerebellar input, the climbing fibers, provide a training signal that, 
thanks to conjunctive LTD on the parallel fiber synapses into Purkinje cells, allows for
the right patterns to be selected. 
Conjuctive LTD (Long-Term Depression) reduces the strength of parallel fiber 
synapses when they happen to be active at the same time as climbing fiber inputs.
Within this general framework, a persistent challenge  comes in 
determining what the right patterns are, and how they are used to improve motor control. 

One common trend for cerebellar models of motor control is to assume that the cerebellum
is involved in providing anticipative corrections to performance errors \cite{manto_consensus_2012},
and that this is done by forming internal models of the controlled objects (Wolpert98,Ebner13).
Forward models take as inputs a command and a current state, returning the consequences of that command,
often in the form of a predicted state.
Inverse models take as their input a desired state and a current state, returning the commands
required to reach the desired state. Adaptive learning in the cerebellum is often assumed to involve using
error signals to learn these types of internal models. It should be noted that some computational
elements (such as adaptive filters), which could be implemented by cerebellar microzones, can in principle
learn to implement either a forward or an inverse model depending on its input/output connections and on 
the nature of its error signal \cite{porrill_adaptive_2013}.

\begin{figure}
  \centering\includegraphics[height=4in]{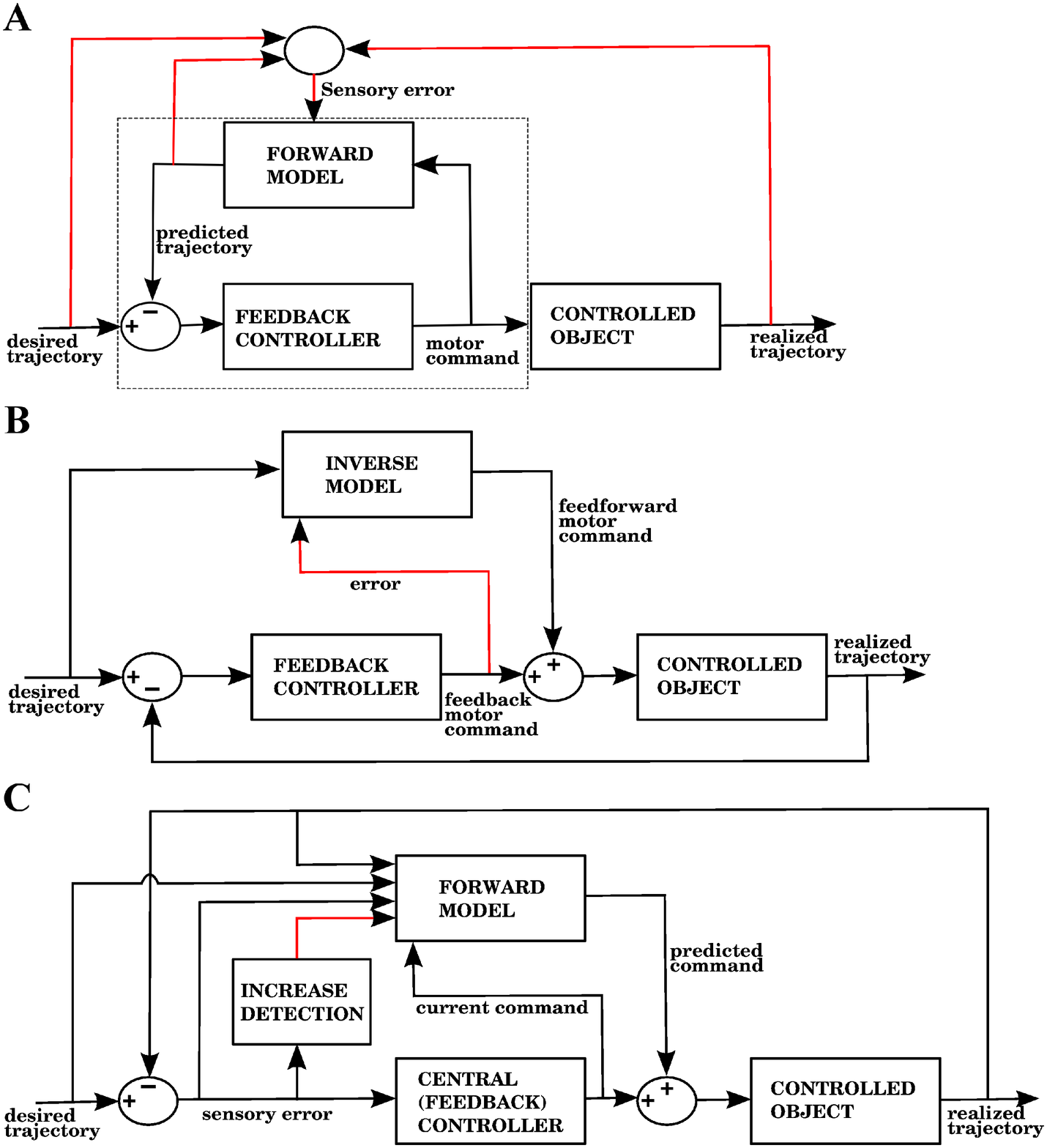}
  \caption{\small A) The recurrent architecture \cite{porrill_recurrent_2007} 
          uses a forward model as the adaptive element in a controller.
	  This forward model learns to predict the response of
	  the controlled object to the motor commands, using an error that considers 
	  the difference between the predicted trajectory and the realized trajectory.
	  Notice that the elements inside the dashed rectangle constitute an adaptive
	  inverse model of the controlled object.
	  Red lines indicate signals used for training of the forward model.
	  Based on figure 1A of \cite{ito_error_2013}.
	  B) Use of an inverse model to improve the performance of a feedback controller
	  using the feedback error learning scheme of \cite{kawato_computational_1992}.
	  The output of
	  the feedback controller is used to approximate the error in the motor command,
	  so the inverse model can be trained. The red line indicates the learning signal.
	  C) A forward model proposed in this paper is used to improve the performance
	  of a central controller. The forward model associates a context consisting of a
	  variety of sensory and motor signals (black arrows entering from the left) with a 
	  command produced by the controller (black arrow entering from below). 
	  The context will be associated with future controller commands whenever the 
	  sensory error increases, indicated by the red line. Notice that while the forward
	  model in panel A predicts the response of the controlled object, the forward
	  model in panel C predicts the response of the central controller.
	  In the Results section, model 1 corresponds to this architecture.
  }
  \label{fig.cb_blocks1}
\end{figure}

The error signal required by a forward model is a sensory error, which consists of the difference
between the desired sensory state (e.g. a hand trajectory) and the perceived sensory state. In
contrast, inverse models require a motor error signal that indicates the difference between
a given command and the command that would have produced the desired outcome.
Figure \ref{fig.cb_blocks1} A,B shows two well known proposed architectures that allow the cerebellum
to use forward and inverse models to reduce performance errors, respectively called the
recurrent architecture, and feedback error learning.
A recent review \cite{ito_error_2013} examined the signal contents of climbing fibers for
different cerebellar circuits, and found that both sensory and motor errors might be present,
bringing the possibility of having both forward and inverse models in the cerebellum.

Inverse models in the cerebellum present some difficulties. The first one is known as the
motor error problem, and consists on the requirement that the climbing fibers carry an
unobservable motor error rather than the observed sensory error signal. This creates
difficulties when applying them to the control of complex plants \cite{porrill_adaptive_2013}. 
A second difficulty is the evidence that simple spikes in Purkinje cells are consistent with a 
forward model, but probably not with an inverse model \cite{ebner_cerebellum_2008}. 
Although climbing fiber may carry information about motor errors, most studies seem to find 
correlations with sensory signals and sensory errors
(e.g. \cite{EkerotGarwiczSchouenborg91,YanagiharaUdo94,GhelarducciItoYagi75,SimpsonBeltonSuhEtAl02,StoneLisberger86,KitazawaKimuraYin98,YanagiharaUdo94}). 

There are two other problems that must be addressed by cerebellar models that form
internal models, whether forward or inverse \cite{porrill_recurrent_2007}.
The distal error problem happens when we use output
errors (such as sensory signals) to train the internal parameters of a neural network.
Backpropagation is a common ---although biologically implausible--- way to deal with this problem.
The nature of the distal error problem is the same one as that of the motor error problem,
since they both are credit assignment problems; in this paper they are used interchangeably.
The redundancy problem happens when a set of commands lead to the same outcome, leading to
incorrect generalizations when that set is non convex. One common setting where the redundancy
problem arises is in reaching. The human arm, including the shoulder and elbow joints has
5 degrees of freedom (without considering shoulder translation), allowing many joint configurations
that place the hand in the same location.

The recurrent architecture of figure \ref{fig.cb_blocks1}A, and the feedback error learning scheme
of figure \ref{fig.cb_blocks1}B are shown here because they present two different ways of addressing 
the motor error and redundancy problems. The recurrent architecture is trained with sensory error, 
so the motor error problem is not an issue; moreover, this architecture receives motor commands as
its input, so it doesn't have to solve the redundancy problem. Feedback error learning 
approximates the motor error by using the output of a feedback controller. The feedback controller
thus acts as a transformation from sensory error into motor error. If the feedback
controller can properly handle redundancy, then so will the inverse model that it trains.

In this paper we propose a new cerebellar architecture that successfully addresses the
motor (or distal) error problem, and the redundancy problem.
This architecture is specified at an abstract level, and consists of descriptions of
the inputs and outputs to cerebellar modules, the content of climbing fiber signals,
and the nature of the computations performed by the cerebellar microzone.

In our architecture, the role of the cerebellum is to provide anticipative corrections
to the commands issued by a central controller (figure \ref{fig.cb_blocks1}C). These
corrections are learned by associating the sensory/motor
context shortly before an error with the corrective response issued by the central 
controller shortly afterwards.
We thus propose that the cerebellar inputs carried by mossy fiber signals consist
of all sensory and motor signals that can be used to predict a future state. The cerebellar
output consists of a predicted set of motor commands similar to a correction issued by
the central controller in the past. The climbing fiber activity rises in response to 
an {\em increase} of an error measure over time, not to instantaneous error values.
Cerebellar microcomplexes act to associate a particular sensory/motor context with a 
response by the central controller happening shortly after an increase in the climbing
fiber activity. This is consistent with many models based on the Marr-Albus-Ito 
framework. If a bank of filters (presumably arising from computations in the granule cell
layer) are placed in the inputs, then this associator becomes functionally similar to
adaptive filter models commonly found in cerebellum literature 
\cite{Fujita82,dean_adaptive-filter_2008}.
Those models usually assume that mossy fiber inputs correlated with climbing
fiber activity cause a decrease in the firing rate of Purkinje cells because of 
conjunctive LTD, leading to an increase in firing rate at the cerebellar or
vestibular nuclei. This could be conceived as associating a particular pattern
of mossy fiber inputs with a response in cerebellar nuclei, with the input filters
giving the system the ability to recognize certain temporal patterns.

We explore the ideas of our cerebellar architecture by implementing it in computational 
and mathematical models of reaching in 3D space. We chose this task because it
presents challenges that should be addressed by cerebellar models, namely
distal learning, redundancy, and timing. In the context 
of reaching, the idea that the cerebellum could function by anticipatively applying the same 
corrections as the central controller raises valid concerns about stability.
We address these concerns by showing that if
the central controller acts like a force always pointing at the target, and whose 
magnitude depends only in the distance between the hand and the target, then
an idealized implementation of our cerebellar architecture will necessarily reduce 
the energy of the system, resulting in smaller amplitude for the oscillations, and 
less angular momentum. The idealized implementation of the architecture thus yields
sufficient conditions for its successful application.
This result is presented in the Supplementary Material.

In addition to our mathematical model, we implemented four computational models 
of a 3D reaching task embodying simple variations of our proposed architecture. 
The central controller in the four models uses an extension of the Equilibrium-point
hypothesis \cite{FeldmanLevin09}, described in the Materials and Methods section. 
The presence of equilibrium points
permits ways of addressing the motor error problem different than using
stored copies of efferent commands from the central controller, and ways
of detecting errors different than visually monitoring the distance between
the hand and the target. Our four models 
thus explore variations of the architecture, in which either the learning
signal or the corrections are generated using proprioceptive signals from
muscles.
For these models the controlled plant is a 4 DOF arm actuated by 11 Hill-type muscles. 
The cerebellar module associates contexts, represented
by radial basis functions in the space of afferent and motor signals with
corrective motor commands. 
These associations between contexts and motor responses happen whenever a learning signal 
is received, which happens when there is an increase of the error.

As mentioned above, we use two types of errors in our computational models.
The first type of error is the distance between the hand and the target, which
proves to be sufficient to obtain predictive corrections. 
By virtue of using the equilibrium-point hypothesis in the central controller
we can alternatively use a second type of error signal generated for individual 
muscles that extend when they should
be contracting. This allows the cerebellum to perform anticipative corrections in a
complex multidimensional task like reaching using learning signals that arise from
1-dimensional systems. This learning mechanism can trivially be extended to serial
cascades of feedback control systems, such as those posited by Perceptual Control 
Theory \cite{powers_feedback:_1973,powers_behavior:_2005} and Threshold Control Theory 
\cite{FeldmanLevin09,LatashLevinScholzEtAl10}, allowing the cerebellum to perform 
corrections at various levels of a hierarchical organization spanning from individual muscle
contractions to complex cognitive operations. We elaborate on this in the Discussion.

\section{Materials and methods}

\subsection{Physical simulation of the arm}

\begin{figure}
  \centering\includegraphics[height=4in]{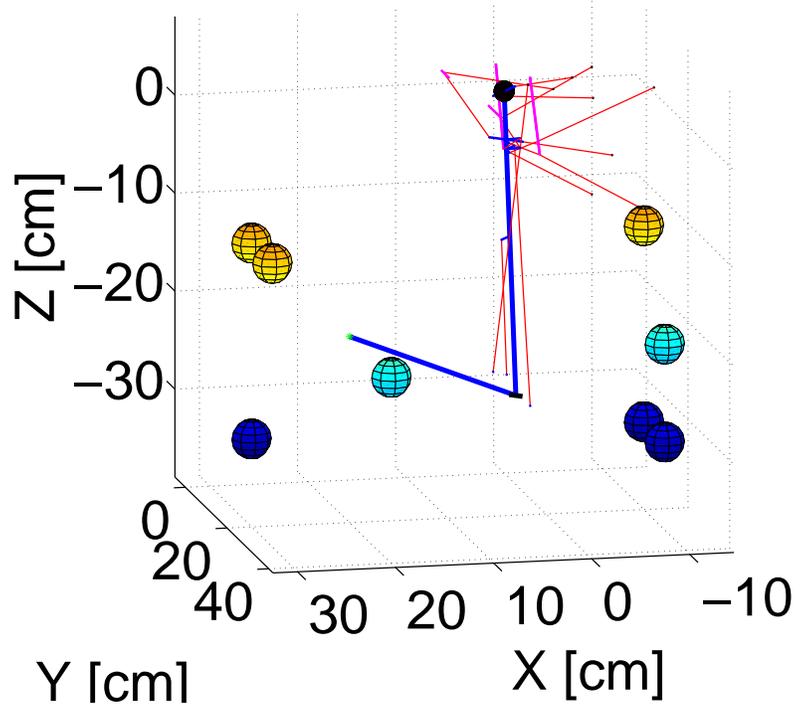}
  \caption{\small Geometry of the arm model. Blue lines represent the upper arm and forearm,
  with the small black sphere representing the shoulder. 
  Red lines represent muscles. Cyan lines are bending lines. The colored spheres (with color 
  representing their position along the Z axis) show the location of the targets used in the
  reaching simulations. The coordinates of these targets are in table \ref{tab:targets}.}
  \label{fig.arm_geom}
\end{figure}

In order to test the principles of our cerebellar model in 3D reaching tasks we created a
detailed mechanical simulation of a human arm. Our arm model contains a shoulder joint
with 3 degrees of rotational freedom, and an elbow joint with one degree of rotational
freedom. Inertia tensors for the arm, forearm, and hand were created assuming a
cylindrical geometry with size and mass typical of human subjects. The actuators consist of 
11 composite muscles that represent the main muscle
groups of the human arm (figure \ref{fig.arm_geom}). Some of these muscles wrap around ``bending lines,''
which are used to model the curved shape of real muscles as they wrap around bones and
other tissue. The force that each muscle produces in response to a stimulus comes from a
Hill-type model used previously with equilibrium point controllers
\cite{GribbleOstrySanguinetiEtAl98}. The mechanical simulation was implemented in
SimMechanics, which is part of the Matlab/Simulink software package
(\verb\http://www.mathworks.com/\), release 2012b. Source code is available from the
first author upon request.

The coordinate for the targets used in our test reaches are shown in table \ref{tab:targets}. \\

\begin{table}
\begin{tabular}{l | c | c | c}
	& X [cm] & Y [cm] & Z [cm] \\ \hline
	Target 1 & -10 & 20 & -30 \\ \hline
	Target 2 & -10 & 20 & -10 \\ \hline
	Target 3 & -10 & 30 & -30 \\ \hline
	Target 4 & -10 & 30 & -20 \\ \hline
	Target 5 &  30 & 20 & -30 \\ \hline
	Target 6 &  30 & 20 & -10 \\ \hline
	Target 7 &  20 & 40 & -20 \\ \hline
	Target 8 &  30 & 30 & -10 
\end{tabular}
\caption{Coordinates used for the targets in the test reaches. The origin is at the
shoulder. The X axis points to the right, the Y axis to the front, and the Z axis 
upwards.}
\label{tab:targets}
\end{table}

\subsection{Central controller}

The central controller we use to perform reaching is a modified version of Threshold 
Control Theory (TCT, \cite{FeldmanLevin09}). TCT is an extension of a biological 
control scheme known as the Equilibrium Point (EP) hypothesis. The lambda version of the
EP hypothesis states that the control signals used in the spinal cord to drive skeletal 
muscles consist of a group of muscle lengths known lambda values. When the length of
a muscle exceeds its lambda value it contracts, so that a set of lambda values will
lead the body (or in our case, the arm) to acquire an equilibrium position. The
muscle lengths at the equilibrium position may or may not be equal to their lambda
values. Also, notice that given a set of lambda values there is a unique position that
the limb will acquire, because the viscoelastic properties of the muscles will lead the
joint to adopt the configuration minimizing its potential energy.
In this paper the control signals arriving at the spinal cord to specify threshold 
lengths for muscle activation are called target lengths. 

Considering that the velocity of a muscle's extension-contraction is represented in 
spindle afferents \cite{Lennerstrand68,LennerstrandThoden68,DimitriouEdin08,DimitriouEdin10},
the argument made for lengths in the EP hypothesis could be modified to hypothesize
threshold velocities being the control signals at the spinal cord level, and threshold 
lengths being used at a higher level, modulating the threshold velocities.
Such a two level control system is inspired by the hierarchical organization found in 
TCT and in Perceptual Control Theory \cite{powers_feedback:_1973,powers_behavior:_2005}, 
and is capable of stabilizing 
oscillations with far more success than pure proportional control. In general, it 
is hard to stabilize movement without velocity information, so this factor has been 
introduced in equilibrium-point controllers \cite{deLussanetSmeetsBrenner02,LiKuanyiAcharya11}. 
As in TCT, we assume that the forces are generated at the level of the spinal cord, 
similarly to the stretch reflex, and we assume a proprioceptive delay of 25 ms.

\begin{figure}
  \centering\includegraphics[height=2in]{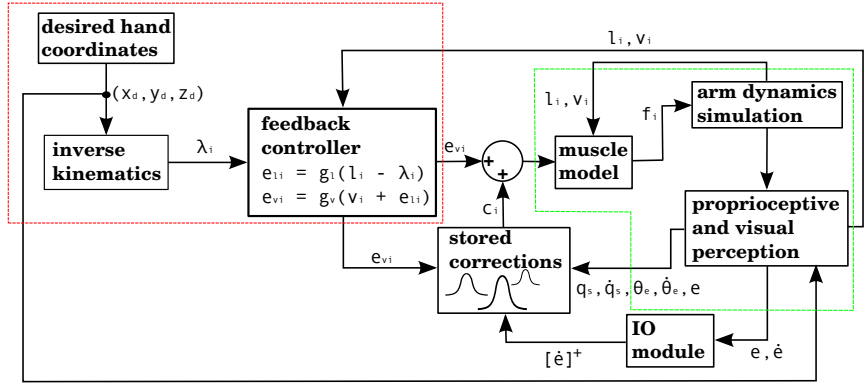}
  \caption{\small Block diagram corresponding to the computational implementation of
	  our architecture in Matlab when using visual errors. 
	  $\lambda_i$ is the target length for muscle $i$.
	  $e_{li}$ and $e_{vi}$ are respectively the length and velocity errors for 
	  the $i$-th muscle. $c_i$ is the cerebellar correction applied to muscle $i$.
	  $f_i$ is the force vector applied on the insertion points of muscle $i$ as
	  a result of its contraction.
	  $l_i$ and $v_i$ are the length and contraction velocity of muscle $i$. 
	  When these signals come directly from the arm dynamics simulation they
	  are not delayed.
	  $q_s$ is a quaternion specifying the orientation of the upper arm. 
	  $\theta_e$ is the angle of elbow flexion. 
          $e$ is the distance between the hand and the target.
	  $[\dot e]^+$ is the positive part of the derivative of $e$. \
	  $l_i,v_i,q_s,\dot{q}_s,\theta_e,\dot{\theta}_e$ are subject to a proprioceptive
	  delay of 25 ms, whereas $e$ and $\dot e$ are subject to a visual delay of 150 ms.
	  The blocks inside the red and green dashed lines are used for the 4 models in 
	  the paper. The elements inside the red dashed square comprise the central controller
	  in figures \ref{fig.comp_model1}, \ref{fig.comp_model2}, \ref{fig.comp_model3}, and
	  \ref{fig.comp_model4}.
	  The blocks surrounded by the green dashed lines constitute the muscle, environment,
	  and parietal cortex blocks in figures \ref{fig.comp_model1}, \ref{fig.comp_model2}, 
	  \ref{fig.comp_model3}, and \ref{fig.comp_model4}.
	  Implementation of the blocks is described in the Material and Methods section.
  }
  \label{fig.cb_blocks0}
\end{figure}

The way our controller guides reaching starts by mapping the Cartesian
coordinates of a target into the muscle lengths that the arm would have with
the hand located at those coordinates. In order to make this mapping one-to-one
we assume that the upper arm performs no rotation. The difference between
the current muscle length and the target muscle length will produce a muscle
stimulation, modulated by the contraction velocity (details in next subsection).
The blocks labeled ``inverse kinematics'' and ``feedback controller'' in figure 
\ref{fig.cb_blocks0} represent the computations of the central controller being described.

\subsubsection{Equations for the central controller}
\label{subsec:eqs}

The central controller performs two tasks in order to reach for a target. The first task is, given the coordinates of the target, to produce the muscle lengths that would result from the hand being at those coordinates. The second task is to contract the muscles so that those target lengths are reached.

The first task (inverse kinematics) requires to map 3D desired hand coordinates into an arm
configuration. The spatial configuration of the arm that leads to hand location
is specified by 3 Euler angles $\alpha, \beta,\gamma$ at the shoulder joint, 
and the elbow angle $\delta$. Our shoulder Euler angles correspond to intrinsic
ZXZ rotations. In order to
create a bijective relation between the 3D hand coordinates and the four arm
angles we set $\gamma = 0$.

For a given target hand position we calculate the angles
$\alpha,\beta,\gamma,\delta$ corresponding to it. Using these angles we
calculate the coordinates of the muscle insertion points, from which their
lengths can be readily produced. When the muscle wraps around a bending line we
first calculate the point of intersection between the muscle and the bending
line. The muscle length in this case comes from the sum of the distances
between the muscle insertion points and the point of intersection with the
bending line.

The formulas used to calculate the angles $\alpha, \beta,\gamma, \delta$ given hand coordinates 
(x,y,z) and the shoulder at the origin are:

\begin{equation}
\alpha = \sin^{-1} \left( \frac{-x}{\sqrt{x^2 + y^2}} \right),
\end{equation}

\begin{equation}
\beta = \cos^{-1} \left( \frac{-z}{\sqrt{x^2 + y^2 + z^2}} \right) -
                \cos^{-1} \left( \frac{x^2 + y^2 + z^2 + L_{arm}^2 - L_{farm}^2 }{2(x^2+y^2+z^2)L_{arm}} \right),
\end{equation}

\begin{equation}
\gamma = 0,
\end{equation}

\begin{equation}
\delta = \pi -  \cos^{-1} \left( \frac{L_{arm}^2 + L_{farm}^2  - (x^2 + y^2 + z^2)}{2 L_{arm}L_{farm}} \right).
\end{equation}

Where $L_{arm}$ and $L_{farm}$ are the lengths of the upper arm and forearm respectively.
If we have the coordinates of a humerus muscle insertion point (as a column vector) at the resting position, 
then we can find the coordinates of that insertion point at the position specified by $\alpha,\beta,\gamma$ 
using the following rotation matrix: 

\begin{equation}
R = \begin{bmatrix} 
\mbox{c}(\alpha)\mbox{c}(\gamma)-\mbox{s}(\alpha)\mbox{c}(\beta)\mbox{s}(\gamma) &
-\mbox{c}(\alpha)\mbox{s}(\gamma)-\mbox{s}(\alpha)\mbox{c}(\beta)\mbox{c}(\gamma) &
\mbox{s}(\alpha)\mbox{s}(\beta) \\
\mbox{s}(\alpha)\mbox{c}(\gamma)+\mbox{c}(\alpha)\mbox{c}(\beta)\mbox{s}(\gamma) &
-\mbox{s}(\alpha)\mbox{s}(\gamma)+\mbox{c}(\alpha)\mbox{c}(\beta)\mbox{c}(\gamma) &
-\mbox{c}(\alpha)\mbox{s}(\beta) \\
\mbox{s}(\beta)\mbox{s}(\gamma) & \mbox{s}(\beta)\mbox{c}(\gamma) & \mbox{c}(\beta)
\end{bmatrix}
\end{equation}

where $\mbox{c}(\cdot) = \cos(\cdot), \ \mbox{s}(\cdot) = \sin(\cdot)$.


The coordinates of insertion points on the forearm at the pose determined by $\alpha,\beta,\gamma,\delta$ are
obtained by first performing the elbow ($\delta$) rotation of the coordinates in the resting position, and 
then performing the shoulder rotation ($\alpha,\beta,\gamma$). Muscle lengths come from
the distance between their insertion points, or between their insertion points and their
intersection with the bending line.
Details on how to determine whether a muscle intersects a bending line can be found in the
function piece5.m, included with the source code. This function also obtains the point of
intersection, which is the point along the bending line that minimizes the muscle length.

Once we have found target equilibrium lengths for the muscles, we must contract them until they adopt those lengths.
To control the muscles we use a simple serial cascade control scheme.
The length error $e_l$ of a muscle is the difference between its current length
$l$ and its equilibrium length $\lambda$. The velocity error $e_v$ is the
difference between the current contraction velocity $v$ (negative when the
muscle contracts), and the length error $e_l$:

\begin{equation}
e_l = g_l (l - \lambda), \quad \quad e_v = g_v (v + e_l).
\end{equation}

The constants $ g_l, g_v $ are gain factors. For all simulations $g_l = 2, g_v = 1$.
The input to the muscles is the positive 
part of the velocity error. This creates a force that tends to contract the muscle 
whenever its length exceeds the equilibrium length, but this force is reduced according 
to the contraction speed. At steady state the muscle lengths may or may not match 
the equilibrium lengths, depending on the forces acting on the arm.
To promote stability the output of the central controller went through a low-pass filter
before being applied to the muscles. Also, to avoid being stuck in equilibria away from the 
target, a small integral component was added, proportional to the time integral of the
central controller's output.

\subsection{Cerebellar model}

The cerebellar model provides motor commands whenever an ``error-prone area'' of state space is entered. Each error-prone area 
consists of a point in state space (its center, or feature vector), and a kernel radius. To each error-prone area there also 
corresponds a ``correction vector,'' specifying which muscles are activated and which are inhibited when the error-prone area 
is entered. At each iteration of the simulation the distance between the currently perceived point in state space and the 
center of each error-prone area is obtained, and each correction vector will be applied depending on this distance, modulated 
by its kernel radius. The kernels used can be exponential or piecewise linear. The action of the cerebellar model
is represented in figure \ref{fig.cb_blocks0} by the block labelled ``stored corrections.''

Learning in the model requires an error signal, which could be visual (such as the one that may be generated in 
posterior parietal cortex \cite{DesmurgetEpsteinTurnerEtAl99}), or could arise from muscle afferents. 
Block diagrams corresponding to the model with the visual and muscle error signals are in figures
\ref{fig.comp_model1},\ref{fig.comp_model2},\ref{fig.comp_model3}, and \ref{fig.comp_model4}.
The visual error signal arrives with a delay of 150 ms. Each time the error increases its magnitude 
(its derivative becomes positive) this increases the probability of complex spikes; for each IO cell, this
probability also depends on the current phase of its subthreshold oscillation (see next subsection). 
Complex spikes generate a new error-prone 
area. The feature vector associated with this area is the state of the system a short time span before the error increased; 
usually this time span will be half the time it takes for the error derivative to go back to zero, plus an amount of time 
comparable to the perceptual delay. For as long as the error derivative is positive, at each iteration we will record the 
efferent signals produced by the central controller, and when the derivative stops being positive we will obtain the average 
of all the recorded efferent signals. The correction is obtained from this average. The muscles are driven by the velocity 
errors, so these are the efferent signals collected during correction period. 
All the kernel radii were equal, so they have no change associated with learning.

Notice that if the error derivative remains positive, more complex spikes will be generated as different 
olivary nucleus cells reach the peak of their subthreshold oscillations. Thus, we have two gain mechanisms for 
a correction: one comes from the magnitude of the error derivative, which will promote a large response 
(and synchronous activity) of complex spikes; the second comes from the amount of time that the error 
derivative remains positive, since more inferior olivary nucleus cells reaching the peak of 
their subthreshold oscillations while this derivative is positive will mean a larger number of complex spikes,
creating error-prone areas along the trajectory of the arm. Performance-wise, it is beneficial to have a
sequence of error-prone areas rather than a single one, since the appropriate correction to apply
will change as the arm moves.

When the new feature vector is too close to a previously stored one, or when we have already stored 
too many feature vectors, then the new feature vector will become ``fused'' with the stored feature 
vector closest to it. When two areas fuse they are both replaced by a new area whose 
feature vector is somewhere along the line joining the feature vectors of its parent 
areas, and likewise for its correction vectors.

\subsubsection{Algorithm for the cerebellum simulations}

We will describe the part of the computational model that deals with the functions of a microcomplex 
(the file CBloop11c.m of the source code). To simplify the exposition, we do not consider the case 
when the maximum number of ``feature vectors'' have been already stored.

The input to the microcomplex model has components that represent error, and afferent/efferent signals. The error component 
consists of the distance between the hand and the target (the visual error), and its derivative (from which complex spikes 
are generated). The afferent information includes a quaternion describing the shoulder joint position, the derivative of 
this quaternion, an angle describing the elbow position, and this angle's derivative. The efferent input is the muscle input 
described in section \ref{subsec:eqs} (consisting of 11 velocity errors), and in addition, the desired shoulder position (expressed 
as a quaternion), and the desired elbow angle. The error and its derivative arrive with a visual delay of 150 ms. The rest 
of the information arrives with a proprioceptive delay of 25 ms.

The output of the microcomplex consists of 11 additional signals that will be added to the muscle inputs.

The algorithm's pseudocode is presented next. An unhandled spike is a complex spike whose ``context'', consisting of the 
afferent/efferent signals and the error briefly before the spike, has not been stored as a ``feature vector''. A ``feature 
vector'' is a context associated with a motor correction. 

 At each step of the simulation:
 \begin{algorithmic}
	\STATE {\bf 1:} Generate complex spikes using the error derivative \\
	{\bf 2:} \IF{there are unhandled spikes} 
   	 \IF{If the error derivative is no longer positive, and the time since the spike doesn't exceed 250 ms}
		\STATE {\bf 2.1.1:} Store the context corresponding to the unhandled spike as a new feature vector
		\STATE {\bf 2.1.2:} Store the motor correction associated with the new feature vector
      \ENDIF
   \ENDIF
   \STATE {\bf 3:} For each feature vector, calculate its distance to the current context, and add its motor correction to the output as a function of that distance  
\end{algorithmic}

In step 2.1.1, the stored feature vector consists of the context as it was  $\tau_v - \tau_p + \frac{t - t_{cs}}{2}$  milliseconds before the complex spike, with  $\tau_v$  being the visual delay,  $\tau_p$  the proprioceptive delay,  $t$  the current time, and  $t_{cs}$  the time when the complex spike arrived. 

In step 2.1.2, the motor correction that gets stored is the average motor input from  $(t_{cs}-\tau_v+\tau_p)$  to  $(t-\tau_v+\tau_p)$.

The output that the microcomplex provides at each simulation step is obtained using radial basis functions. The distance between the current context and each feature vector is calculated, and those distances are normalized. The contribution of each feature vector to the output is its corrective motor action scaled by an exponential kernel using that normalized distance.
Let $f(i)$ be the  $i$-th feature vector, and  $w(i)$ its associated correction. Let  $v$ denote the vector with the current context information. We first obtain a distance vector  $D$, whose components are: $D(i) = \| f(i) - v \|^ 2 $.

The distance vector is normalized as $D_N = (\sqrt{M_F}/ \| D \|)D ,$ where $M_F$ is the maximum number of feature vectors allowed. The contribution of feature $i$ to the output is $F(i) = w(i) e^{\gamma D_N(i)}$, with $\gamma$ specifying the kernel radius.

\subsection{Inferior olivary module}

The process of generating complex spikes when using the visual error is explained 
next. By ``complex spike'' we mean a signal indicating that a correction should be stored.
There are  $N$ inferior olivary nucleus cells, from which $N_3$ are assumed to oscillate 
at 3 Hz, and $N_7$ are assumed to oscillate at 7 Hz. The phases of both cell subpopulations 
are uniformly distributed so as to occupy the whole range $[0, 2 \pi]$ in the equation 
below. Let $\phi(i)$ denote the phase of cell $i$, and $\alpha(i)$ denote its 
angular frequency. The probability to spike at time $t$ for cell $i$ is calculated as:

\begin{equation}
P_{CS}^i(t) = p \frac{ \cos[\alpha(i)(t - \phi(i))] + 1}{(1 + e^{(5 - E)})(1 + e^{(30 - 15[E']^+)})}
\label{eq:pcs}
\end{equation}

Where $p$ is a constant parameter, $E$ is the visual error, and  $[E']^+$ is the positive part of its derivative.
At each step of the simulation a random number between 0 and 1 is generated for each cell. 
If that number is smaller than  $P_{CS}^i$, and the cell $i$ has not spiked in the 
last 200 ms, then a complex spike is generated.

Complex spikes are less likely to be generated when the error is small. When the hand is close
to the target it is likely that it oscillates around it. Generating cerebellar corrections in this situation
could be counterproductive, as the angle between the hand and the target changes rapidly, and so do the
required corrections. In our idealized cerebellum (see Supplemenatry Material) there are conditions
ensuring that no corrections are created when the angle between the hand and the target has changed too much.
Since there is no obvious biological way to measure the angle between the hand and the target, we just
avoid generating corrections when the hand is close to the target.
Another mechanism present in our computational simulations to deal with this problem is
that no corrections are stored if the time between the complex spike and the time when the error stops 
increasing is more than 250 ms.

Generating complex spikes when using the proprioceptive error follows a simpler procedure. For each 
muscle three conditions must be satisfied for a ``complex spike'' to be generated: 1) its length $l$ is 
increasing, 2) $l$ is longer than it's target value $\lambda$, and 3) no complex spikes have been 
generated for that muscle in the last 200 ms. A variation described in the Results section adds a
fourth condition: 4) the visual error must be increasing ($E' > 0$).

\subsection{Generating corrective muscle activity}

In this paper there are three different methods to determine the corrective motor commands that become
associated with points of state space where the error increases.

\begin{figure}
  \centering\includegraphics[height=4in]{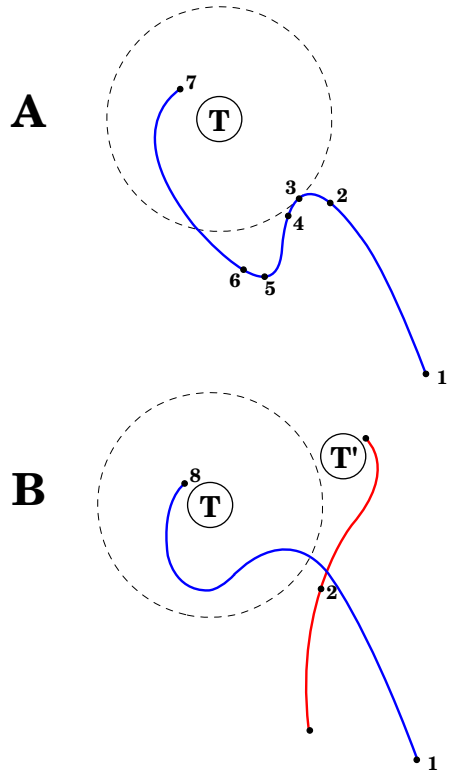}
  \caption{\small Correcting reaching errors. A) Schematic trajectory of the hand as 
	  it reaches for target T in 2
	  dimensions. Seven points of the trajectory are illustrated, corresponding to
	  seven important points in time with different afferent/efferent contexts. {\bf
	  1.} Initial position of the hand. {\bf 2.} The context at this point will be
	  associated with the correction. {\bf 3.} The error begins to increase. {\bf 4.}
	  Complex spikes reach the cerebellar cortex in response to the error increase.
	  {\bf 5.} The error is no longer increasing. {\bf 6.} The context at point 2
	  becomes associated with a correction, which could consist of the mean efferent
	  activity (roughly) between points 3 and 5. {\bf 7.} Final hand position. B)
  After the correction in A) is learned, and the same reach is attempted, the trajectory
  will be modified upon approaching point 2, with the correction being applied
  anticipatively (blue line). Notice that a different trajectory (red line) that passes
  through the spatial location of point 2 may not elicit the correction learned in A).
  This is because the correction is applied when its associated context is near to the
  current context (which is a point in state space); those contexts contain velocities,
  efferent activity, and target location in addition to the arm's spatial configuration.}
  \label{fig.cb_reaching}
\end{figure}

The first method, in model 1 of the Results section, is used with visual errors. 
The corrective commmand consists of the average efferent 
commands produced from the point when the error started to increase until the error stopped increasing 
(points 3 and 5 in figure \ref{fig.cb_reaching}A).

The second method, in models 2 and 3, is used with proprioceptive errors. 
If a complex spike is generated for a muscle, 
the corrective command is simply a slight contraction of that same muscle.

The third method is used with visual errors, and is applied in model 4. The corrective command for muscle $i$ 
will be proportional to the product $ c_i = [<l_i>-\lambda_i]^+ [\dot l_i]^+$, where $l_i$ is the length
of muscle $i$, $<l_i>$ is the average of that length through a brief period before the error stopped
increasing (e.g. a brief period between points 3 and 5 in figure \ref{fig.cb_reaching}A, $\lambda_i$ is
the target length for muscle $i$, $\dot l_i$ is the derivative of the length, and $[\cdot]^+$ returns 
the positive part (and zero otherwise).

\section{Results}

\subsection{Implementing the architecture in a reaching task}

We hypothesize that the role of the cerebellum in motor control is to associate 
afferent and efferent contexts with movement corrections produced by a central controller;
in the case of reaching the controller involves the cortex, basal ganglia, brainstem, and spinal
cord.
The role of the central controller is to reduce error, and the role of the cerebellum 
is to anticipatively apply the corrections of the central controller. How this could
happen for the case of reaching is described in Figure~\ref{fig.cb_reaching}. 
Before an incorrect motion is made 
(moving the hand away from the target), the mossy fibers reaching the granule layer 
have afferent and efferent information that could predict when this error will occur.  When the 
error does increase during a reach, this is indicated by complex spikes, while the 
central motor controller is acting to correct the error.  The cerebellum associates 
the afferent and efferent information of granule cells shortly before the increase 
in error with the motor actions required to correct it, using climbing fiber activity 
as the training signal.
The corrective motor actions can be those that the central motor controller produces in 
order to stop the error from increasing, which come shortly after the onset of error 
increase; thus the cerebellum doesn't have to obtain those actions itself,
it can merely remember what the  central controller did. This idea is related to
Fujita's feed-forward associative learning model \cite{Fujita05}. Other ways to
obtain the corrective motor actions are described in the models below.

As mentioned in the Introduction, we created mathematical and computational models 
implementing these ideas. The mathematical model and the results of its analysis are
described in the Discussion. The full mathematical treatment is in the Supplementary
Material. The elements of the computational models are described in the Materials and Methods
section. In the remainder of the Results we present the outcome of simulations using
four computational models with basic variations of our cerebellar architecture. 
All these computational models use the same central controller and the 
same arm and muscle models.

The physical simulation of the arm used for this study used no friction at the joints.
The muscles had limited viscoelastic properties and the control signals had low
gain. Under these conditions, the arm under the action of the central controller
alone tended to place its distal end at the target slowly
(in around 1.5 seconds) and with some oscillations, even in the absence of gravity forces.
Introducing a 25 ms proprioceptive delay resulted in larger oscillations, and the hand
no longer reached the target with arbitrary accuracy, but would instead oscillate 
around it in a non periodic fashion. Moreover, certain positions of the target would cause
the arm to become unstable, leading to chaotic flailing.

To test that the cerebellar corrections could gradually reduce the error as learning progressed
through successive reaches, we selected 8 target locations and simulated 8 successive reaches
to each target. From these 8 targets one of them (target 2) produced instability of the
arm when no cerebellar corrections were applied. The same 8 targets were used for the
four models presented here. Figure \ref{fig.arm_geom} presents a visualization of the arm's
geometry, and of the 8 targets.

\subsection{Simulation results}
\label{subseq:sim_res}
\subsubsection{Model 1: visual errors, efferent copies to generate corrections.}

\begin{figure}
  \centering\includegraphics[width=5in]{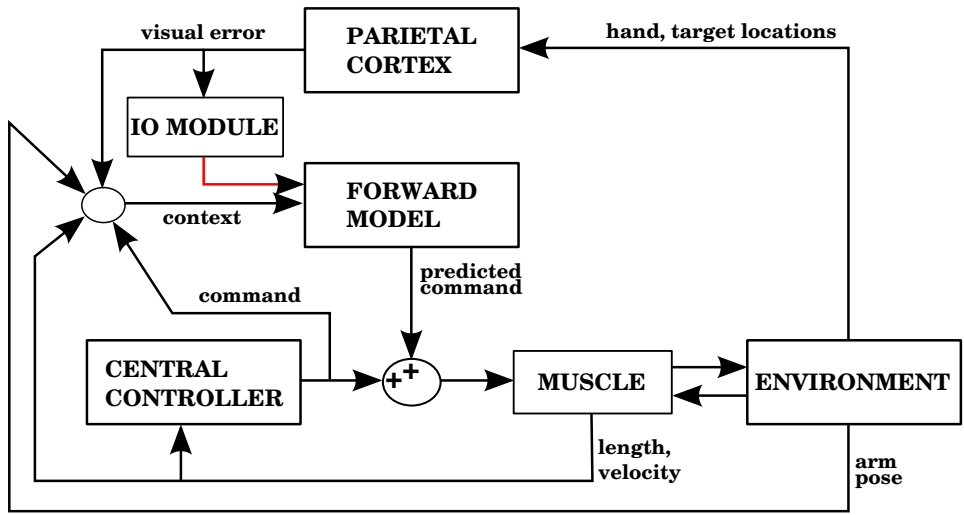}
  \caption{\small Computational model with the visual error signal, and a corrective command that
	  is obtained from the efferent commands produced by the central controller (model 1 in the text). 
	  This is the same model depicted in figure \ref{fig.cb_blocks0}, but at a slightly higher level
	  of description.
	  The error (assumed here to be obtained in parietal cortex) consists 
	  of the distance between the hand and the target, and increases of this error cause the forward
	  model to associate the context with a correction. The learning signal, produced when the error
	  increases, is denoted by the red line. The forward model corresponds to the stored corrections
	  in figure \ref{fig.cb_blocks0}, and the environment corresponds to the arm dynamics simulation.
  }
  \label{fig.comp_model1}
\end{figure}

We first considered the case when complex spikes were generated when the distance between the hand and 
the target increased,
according to equation \ref{eq:pcs}. The corrective muscle commands were proportional to the average of the 
efferent commands produced between the onset of error increase and the time when the error no longer 
increased (the period between points 3 and 5 in figure \ref{fig.cb_reaching}). Figure \ref{fig.comp_model1}
presents a block diagram indicating the signals and modules involved in this model.

\begin{figure}
  \centering\includegraphics[height=4in]{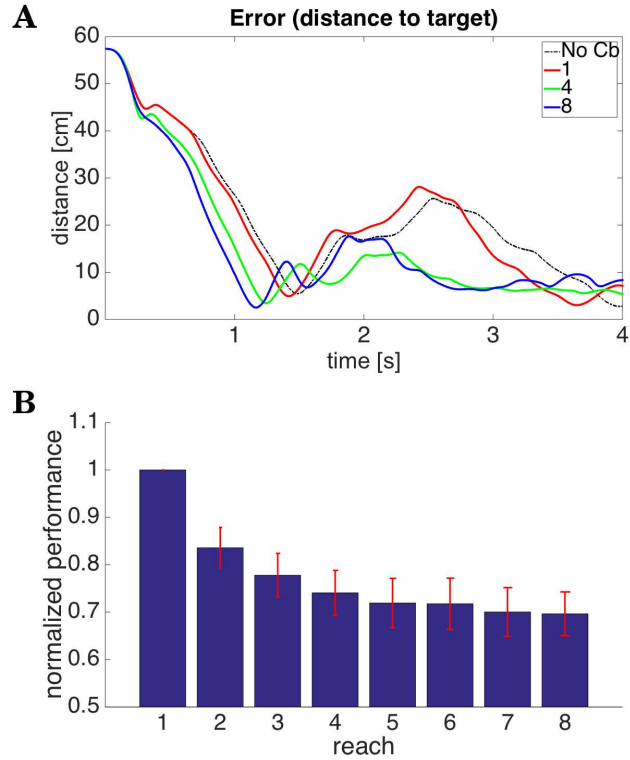}
  \caption{\small Results for Model 1.
	  A) Distance between hand and target through 4 seconds of simulation time for the first, fourth, 
	  and eighth reaches to target 7. The cerebellar system was trained using the distance between the hand and the
	  target as the error, and the target had coordinates X=20 cm,Y=40 cm,Z=-20 cm. 
	  The dashed line, labeled ``No Cb'', shows the error when the arm was actuated by the
	  central controller exclusively.
	  Notice how the first reach (red line) is slower, and oscillates away from the target 
	  after approaching it. This is significantly improved on the eighth reach (blue line).
	  B) Integral of the distance between the hand and the target during the 4 seconds of simulation for the 8 
	  successive reaches. Each bar corresponds to the value obtained from averaging this performance measure 
          across the 8 targets. The bars were normalized by dividing between the value for the first reach. For
	  each bar its standard error measure \ ($S.D./\sqrt{8}$) is shown using the red lines at its upper edge.}
  \label{fig.reaches_c}
\end{figure}

Panel A of figure \ref{fig.reaches_c} shows the evolution through time of the distance between the hand 
and the target in the 1st, 4th, and 8th reaches towards a representative target. To measure the success 
of a reach we obtained the time integral of the distance between hand and target through the 4 seconds
of simulation for each reach. Smaller values of
this performance measure indicate a faster, more accurate reach. Panel B of figure \ref{fig.reaches_c} 
shows our performance measure for each of the 8 successive reaches, averaged over the 8 targets.

Figure \ref{fig.reaches_c} shows that on average the performance 
increases through successive reaches. The error may not decrease monotonically, however, 
since the correction learned in the last trial may put the system in a new region of state space where 
new errors can arise within the time of the simulation. Eventually, however, the
hand comes close to monotonically approaching the target. The instability present in the
second target dissappeared on the second reach.

Although this model improves the performance of the reach,
it can't be considered biologically plausible unless we understand how the outputs at 
the deep cerebellear nucleus could become associated with the corrections they 
presumably apply. Basically, the problem is that if all microcomplexes 
receive the same learning signal (increase in visual error), then all the DCN populations
will learn the same the same response, and the arm would express all possible
corrections upon entering an error-prone area of state space. 
In the Discussion we elaborate on this. In the rest of the Results section we 
present 3 alternative models were the corrections to be applied are not learned
from efferent copies of the commands to the arm, but from proprioceptive signals.

\subsubsection{Model 2: proprioceptive errors, individual muscle corrective signals.}

\begin{figure}
  \centering\includegraphics[width=5in]{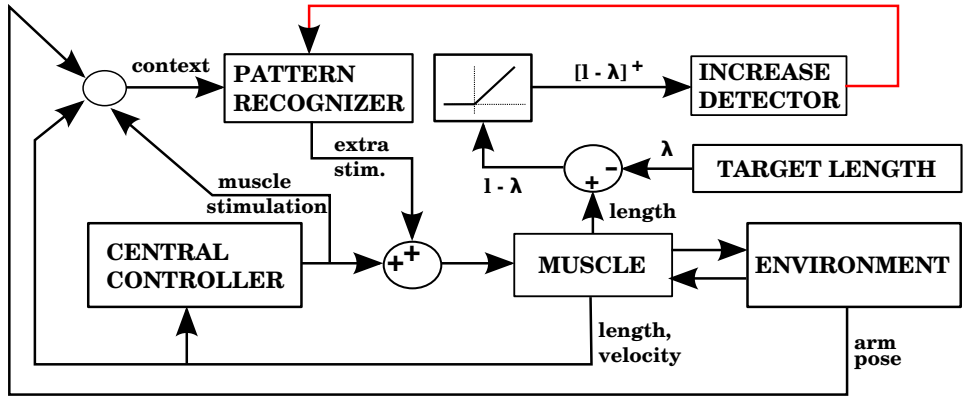}
  \caption{\small Model with the proprioceptive error signal, and a corrective command that is simply a
	  contraction of the muscle that produced the error signal (model 2 in the text). 
	  The error is the muscle length $l$ minus the target length $\lambda$. This target length
	  comes from the central controller. When $l - \lambda$ is positive, increases of this error in 
	  a particular context will cause the pattern recognizer to apply an anticipative contraction 
	  when that context arises. The pattern recognizer corresponds to the block of stored corrections
	  in figure \ref{fig.cb_blocks0}, and the increase detector corresponds to the IO module.
  }
  \label{fig.comp_model2}
\end{figure}

Using the equilibrium point hypothesis in the central controller has the distinct 
advantage that we know the lengths at which the muscle stops contracting (called
target lengths in this paper). A simple way to detect errors could be to monitor
when a muscle is longer than its target length, but is nevertheless elongating.
A simple way to correct that error is to contract that muscle a bit more. The
multidimensional task of applying corrections during 3D reaching is thus 
reduced to a group of one dimensional tasks corresponding to individual
muscle groups. Figure \ref{fig.comp_model2} shows a block diagram implementing
these ideas as done in our second model.

\begin{figure}
  \centering\includegraphics[height=4in]{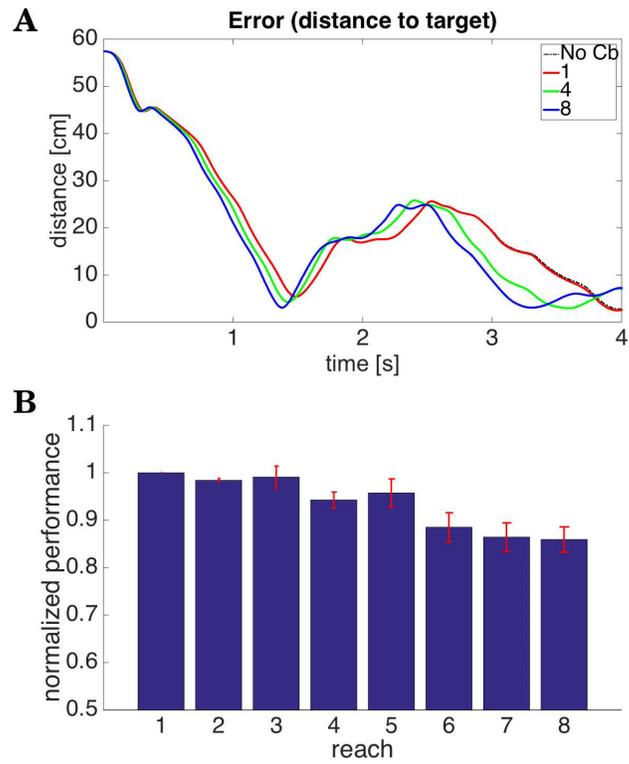}
  \caption{\small Results for Model 2. The
	  cerebellar system was trained using an error signal produced when muscles
          became larger than their target value. 
	  A,B) Refer to figure \ref{fig.reaches_c} for interpretation.}
  \label{fig.reaches_d}
\end{figure}

Figure \ref{fig.reaches_d} shows the results of using a model where the errors
are detected and corrected at the level of individual composite muscles, as
just described. It can be observed that improvement is slower than in the case
of the previous model. For example, the instability of the second target only dissappeared
during the sixth reach (not shown). In our simulations of model 2 the cerebellar corrections 
could lead to instability unless we use small kernel radii and small amplitude for the
corrections. A possible reason for this is that our central controller does not specify an optimal
temporal sequence of muscle contractions, but instead specifies a static set of target lengths. The
trajectory of muscle lengths that leads the hand in a straight line towards the target may not
have those lengths monotonically approaching the target lengths. On the other hand, our system
generates an error signal whenever that approach is non monotonic. This inconsistency is the price
of using one-dimensional signals to approach an error that arises from the nonlinear interaction of
several independent variables. The next model uses a simple approach to try to overcome
this problem.

\subsubsection{Model 3: proprioceptive errors with visual error constraint, individual muscle corrective signals.}

\begin{figure}
  \centering\includegraphics[width=5in]{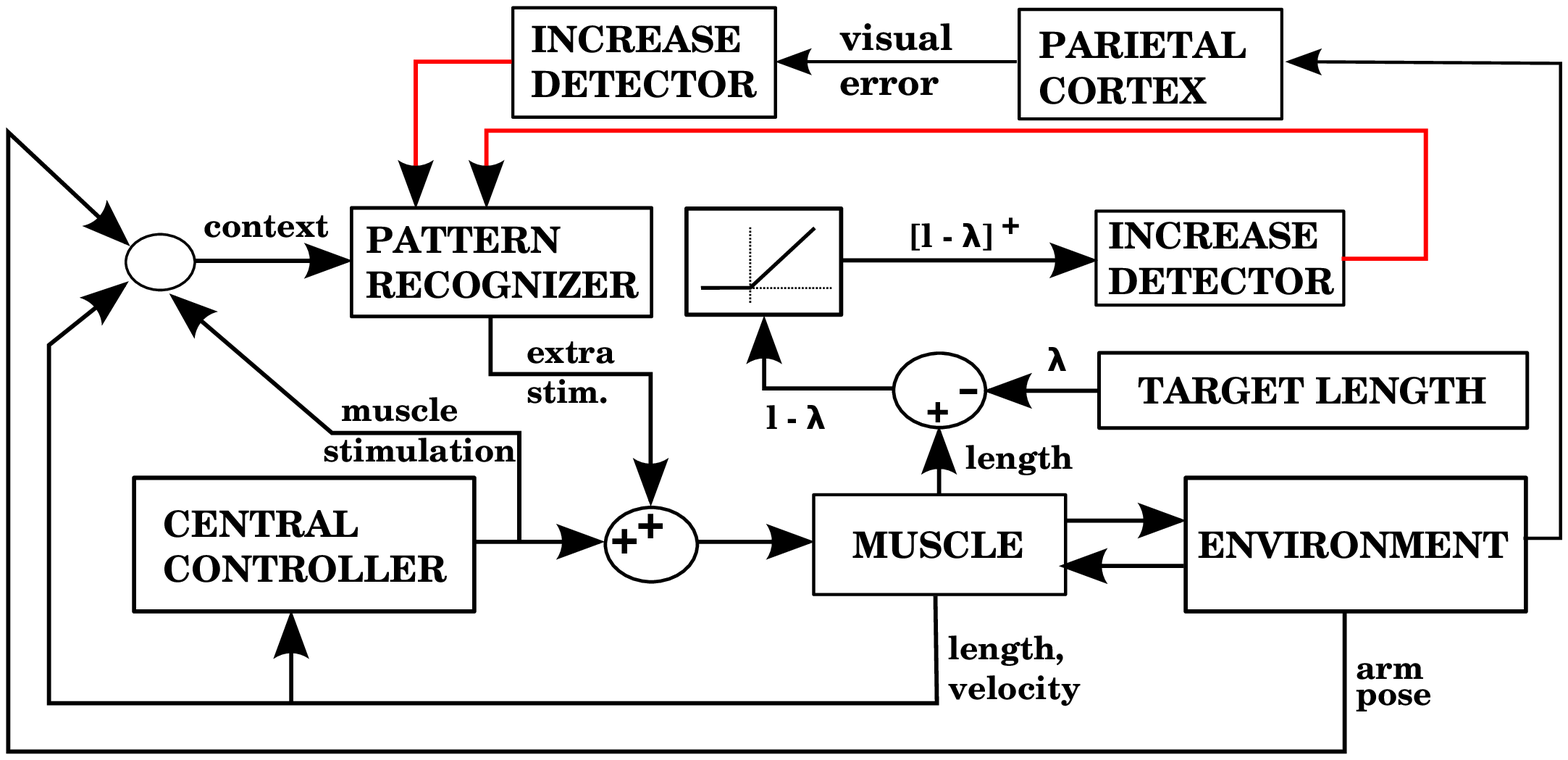}
  \caption{\small Model with the proprioceptive error signal, a visual error constraint, 
	  and a corrective command that is simply a
	  contraction of the muscle that produced the error signal (model 3 in the text). 
	  Notice that this is similar to the model in figure \ref{fig.comp_model2}, but 
	  we have an additional learning signal entering the pattern recognizer. This additional
	  signal ensures that corrections are stored only when the visual error is increasing.
  }
  \label{fig.comp_model3}
\end{figure}

In the previous model the gain of the corrections and their area of application in state
space had to remain small because there can be some inconsistency between the error signals
from individual muscles and the visual error. A muscle continuing to elongate past its
target value does not imply that contracting it will bring the hand closer to the target.
A simple way to address this is to add the necessary condition that if a correction is
to be stored, the visual error should be increasing. Corrective signals will thus arise when the
muscle is elongating beyond its target length, and the hand is getting away from the target.
In this way, even if the muscle lengths are getting away from their target values, 
no corrections will be stored when the hand is approaching the target monotonically. 
Figure \ref{fig.comp_model3} shows how the architecture of model 2 is augmented
with visual errors in order to produce model 3.

\begin{figure}
  \centering\includegraphics[height=4in]{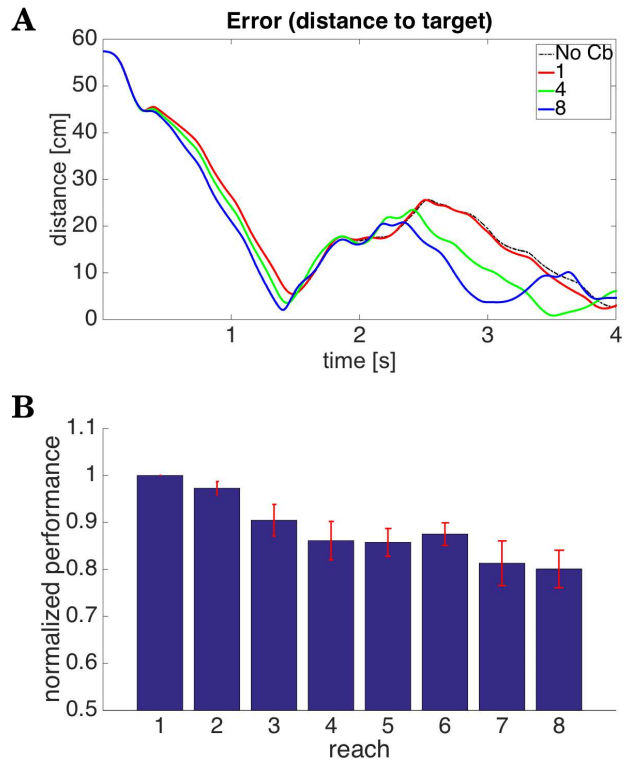}
  \caption{\small Results for Model 3.
          The cerebellar system was trained using an error signal produced when muscles
  became larger than their target value, with the additional constraint that the error
  (distance between hand and target) had to be increasing.
	  A,B) Refer to figure \ref{fig.reaches_c} for interpretation.}
  \label{fig.reaches_d2}
\end{figure}

Figure \ref{fig.reaches_d2} shows the results of using a such a model. Using the additional
constraint permits larger gains in the corrections and larger kernel radii than those used
in model 2. This is reflected by a larger increase in performance. This increase, however,
is still not as good as that seen in model 1. The visual error is what we really want to
reduce, and there is a limit to how much this can be done when the error signals are triggered
at the level of muscles, as the visual error and the proprioceptive error are not entirely
equivalent. This is addressed by the next model.

\subsubsection{Model 4: visual errors, proprioceptive corrective signals.}

As discussed above, visual errors are the most appropriate to improve performance, so
in this model we use them, just as in model 1. Unlike model 1, we don't use the
commands from the central controller in order to create the corrections. We must then
find a way to solve the motor error problem without the central controller. A way 
to do this is to create corrections similar to the signals that indicated error
increase in models 2 and 3.

\begin{figure}
  \centering\includegraphics[height=4in]{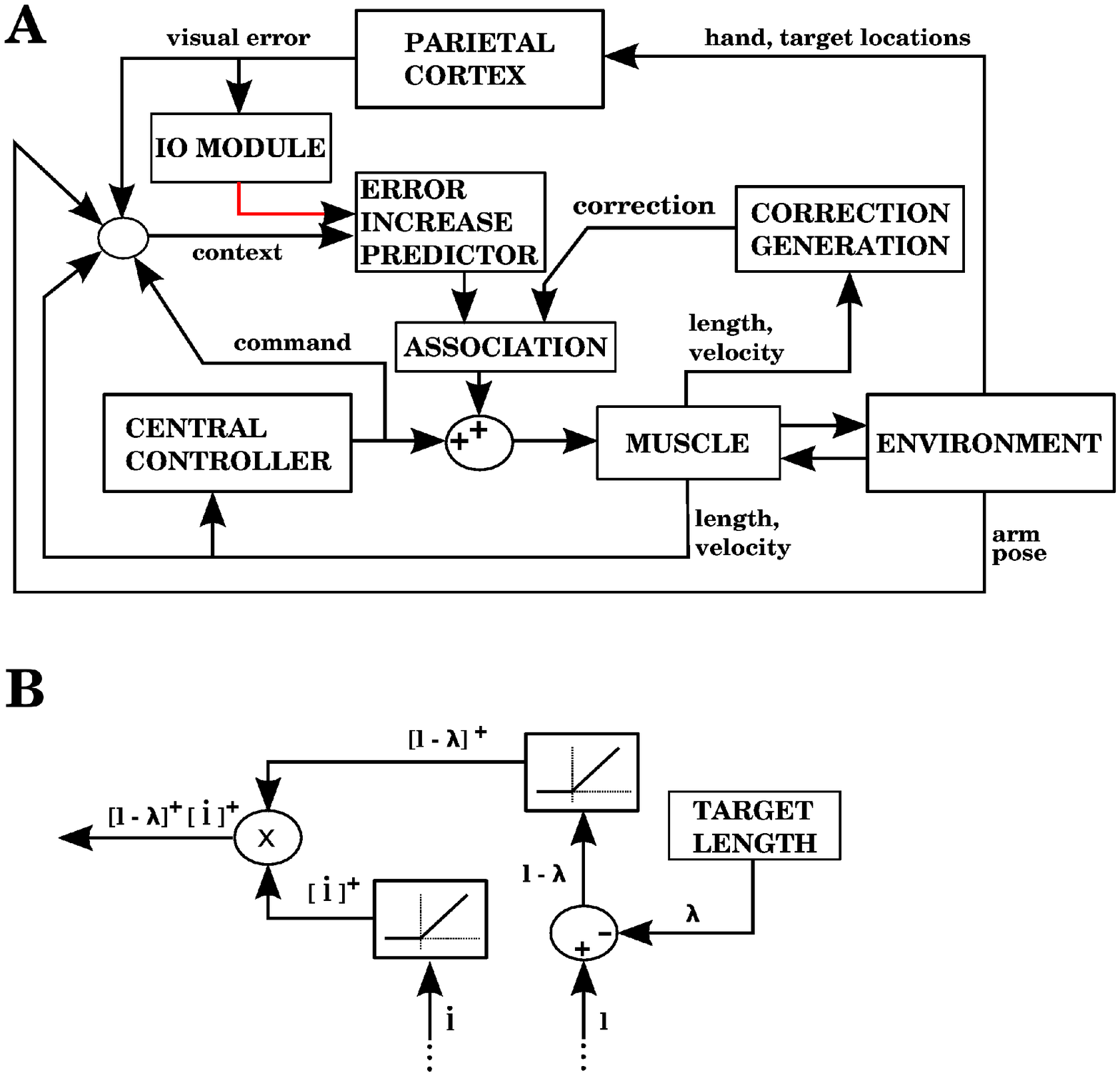}
  \caption{\small Model with visual errors and proprioceptive error signals (model 4 in the text).
	  A) The visual error signal used by this model is the same one as in model 1, but 
	  unlike model 1, the correction associated with an error is not a copy of a command 
	  from the central controller. In this case, the correction is generated from
	  proprioceptive information (muscle length and contraction velocity) in the block
	  labeled as ``CORRECTION GENERATION'' (expanded in panel B). 
	  This correction is to be applied when the 
	  error is predicted to increase. In the block labeled ``ASSOCIATION'' a signal
	  predicting the onset of error increase becomes associated with the correction, so
	  that when the increase in error is predicted the correction is applied.
	  B) The computations performed in the ``CORRECTION GENERATION'' block of panel A.
	  For each muscle, its length $l$ and contraction velocity $\dot l$ are received, along
	  with a target length $\lambda$. The correction consists of the product between the
	  positive parts of $l-\lambda$ and $\dot l$.
  }
  \label{fig.comp_model4}
\end{figure}

Model 4 generates error signals (complex spikes) when the hand is getting away from
the target according to equation \ref{eq:pcs}, just like model 1.
Figure \ref{fig.comp_model4}A shows the signals and modules implied by model 4. 
For each muscle, the correction associated with an error signal is proportional
to two factors: how much longer the muscle is than its target value, and how fast its
length is increasing (figure \ref{fig.comp_model4}B). The block that associates contexts with
predicted increases in error (labeled ``ERROR INCREASE PREDICTOR'') is identified with
the cerebellum, while the ``CORRECTION GENERATION'' module is identified with
muscle afferents and spinal cord neurons. We assume that the predictions of error
increase from the cerebellum become associated with the corrections generated
at the level of the spinal cord. This is elaborated in the Discussion.

\begin{figure}
  \centering\includegraphics[height=4in]{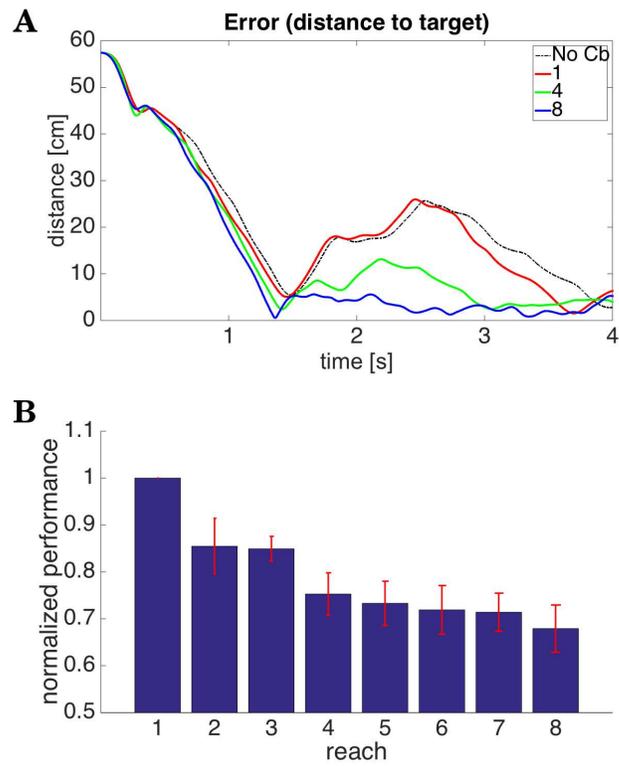}
  \caption{\small Results for Model 4.
          The cerebellar system was trained using the same error signal as in model 1
	  (figure \ref{fig.reaches_c}), but the corrective commands were produced from
	  muscle proprioceptive signals.
	  A,B) Refer to figure \ref{fig.reaches_c} for interpretation.}
  \label{fig.reaches_cd}
\end{figure}

Figure \ref{fig.reaches_cd} shows the performance of model 4. It can be seen that 
the error reduction is comparable to that of model 1, but using a novel solution to
the motor error problem based on the assumption that the muscle is controlled through
an equilibrium length.

\section{Discussion}

As research on the cerebellum continues, it becomes increasingly clear that although cerebellar microzones
have a uniform architecture, the role they play in various systems can be different 
depending on their input and output connections. For example, cerebellar microzones could
implement either forward or inverse models \cite{ito_error_2013,popa_predictive_2012,porrill_adaptive_2013}.
Cerebellar architectures such as feedback-error learning 
(\cite{kawato_computational_1992}, figure \ref{fig.cb_blocks1}B)
and the recurrent architecture (\cite{porrill_recurrent_2007}, figure \ref{fig.cb_blocks1}A)
specify the connectivity of cerebellar microzones, and the computational role they would
therefore play.

We have presented an architecture in which the cerebellum reduces errors associated with climbing fiber activity
when that activity arises from the increase in some error measure. Instead of assuming that 
complex spikes encode the magnitude of some performance error, we have assumed that they are
generated when the derivative of the error becomes positive. This leads to a sparse code that
generates a forward model for anticipative corrections. This forward model exists only in locations 
of state space where the error is prone to increase, and predicts a corrective command,
not the output of the controlled object. Although we have assumed that the central controller uses
closed-loop feedback, this is not necessary. Our model has the potential to explain the presence of predictive
and feedback performance errors in Purkinje cell simple spikes \cite{popa_predictive_2012,popa_purkinje_2013},
the correlation of complex spikes with both sensory and motor events \cite{ito_error_2013}, the sparsity of 
complex spikes, and as discussed below, the role of the cerebellum in nonmotor operations 
\cite{ito_control_2008,buckner_cerebellum_2013,koziol_consensus_2014,popa_cerebellum_2014}.
A possible way to discard our architecture in a particular system is when errors that are not 
increasing or changing elicit a sustained complex spike response. By ``errors'', we refer not only 
to performance errors, but in general to signals that merit a behavioral response, such as an
unexpected perturbation, or a potentially nociceptive stimulus.

We have explored our architecture in the context of reaching in 3D space. 
In addition to the mathematical treatment described below, we showed that the equilibrium
point hypothesis gives our architecture the ability to solve the motor error problem in a
novel way, using proprioceptive muscle signals (models 2,3, and 4). The computational models presented in
this paper show that we can provide predictive control without the need to predict the 
kinematic or dynamic state variables of the controlled plant. Moreover, a signal which very 
loosely represented the positive part of the error derivative is sufficient to train our 
predictive controller. The type of corrections that our model cerebellum provides tend to avoid 
episodes where the hand gets away from the target; this is important when using a controller 
based on the lambda model of the equilibrium-point hypothesis \cite{FeldmanLevin09}. 
A controller that only specifies a set of target muscle lengths (and not a trajectory of such 
lengths) may produce reaches by simultaneously contracting all 
the muscles whose lengths are longer than their desired lengths. 
This, in general, will not result in a straight-line reach. What the cerebellar controller 
does is to modify the activity of antagonist muscles at different points of the trajectory 
so that the hand monotonically approaches its target, producing a reach that is closer to a straight line.

All four models in this paper avoid or solve the redundancy problem. In section 
2.5 three ways of generating corrective motor commands were described. When the 
corrective output is generated from an efferent copy of the central controller 
(model 1), the redundancy problem is avoided, as it is assummed that this is handled 
by the central controller (the recurrent architecture avoids the redundancy problem 
in a similar manner). For the two other ways of generating corrective commands 
(in models 2,3,4), the redundancy problem is solved as soon as equilibrium lengths 
are given. Notice that equilibrium lengths determine the final position of the arm 
uniquely, as the viscoelastic properties of muscles lead the arm towards a configuration 
of minimal potential energy.

\subsection{The mathematical model}

In our mathematical model the hand is considered to be a point mass, and
the central controller applies a force applied to this mass, always
pointing to the origin, which is considered to be target.
This constitutes a central force system, and as in the case of planetery
motion under gravity forces it will tend to produce elliptical trajectories
around the origin.

We modelled the ``cerebellum'' as a system that would apply impulsive forces
to the point mass whenever particular regions of state space were entered,
and proceeded to prove that such a 
cerebellum will continue to reduce the angular momentum in the trajectory
until it either gets close enough to the target, or until it becomes circular.
Circular trajectories do not ellicit cerebellar corrections because the error
signal (distance between the hand and the target) does not increase.
This is a shortcoming of generating learning signals only when the error 
increases. 

The crucial part of this mathematical treatment is specifying when
cerebellar corrections will be created, and for each cerebellar correction
what will be the impulse vector associated with it. The cerebellar 
controller is characterized by three numbers: a speed threshold, a
distance threshold, and a gain. A cerebellar correction
is created whenever two conditions are met: the error begins increasing
faster than the speed threshold, and it grows beyond the distance
threshold.

The impulse associated with a correction is obtained by
integrating the central controller's force, from the time when the error
began to increase, until a stop time is reached; this is then multiplied
by the gain. Specifying the integration stop time
correctly is very important, and in our model we obtain it as the largest
time when three conditions are all satisfied, namely: 1) the error is still
increasing faster than the speed threshold, 2) the mass hasn't rotated around
the origin more than $\pi/2$ radians, 3) the corrective impulse is not strong
enough to reverse the radial velocity of the point mass. The first condition
ensures that we only integrate forces that are contributing to stopping the
error increase. The third condition exists so the corrective impulse is not
strong enough to reverse the velocity of the mass, potentially bringing
instability.

The second condition for the stop time ensures that the impulse vector roughly points in the
opposite direction of the error's velocity vector. This
condition is akin to the strictly positive real (SPR) condition of adaptive
filter models \cite{porrill_adaptive_2013}. The SPR condition states that the error signal
used to train the filter should not have a phase shift of more than
90 degrees at any frequency with respect to the actual error signal. 
In other words, the
SPR condition states that the used error signal should be positively
correlated to the error, whereas our second condition for the integration stop time
states that the corrective signal should be negatively correlated with
the increase in error.  In the next subsection the subthreshold oscillation of 
inferior olivary nucleus cells is linked to the second condition for the 
integration stop time.

The mathematical treatment of our model points to several potential
shortcomings implied in the three conditions for the integration stop
time. These shortcomings are only strengthened by the fact that the arm
does not exactly act as a central force on the hand. The ability of the
cerebellar corrections to generalize properly to points in a ball 
surrounding an original correction point depends on how much the angle
between the error's velocity and the corrective impulse change inside
that ball. The arm exerting forces that don't point towards the target
could reduce its negative correlation with the error velocity. This is a
reason why the computational simulations in this paper (particularly
model 1) are an important validation of our mathematical ideas.

\subsection{The contents of climbing fiber activity}

What the climbing fibers (CF) encode is still a contentious issue, and different assumptions
lead to different models of cerebellar function. One set of assumptions is that the
CF activity encodes performance errors involving the neuronal circuits of the PCs
receiving those CFs. CF activity has indeed been found to be related to performance
errors and unpredicted perturbations 
\cite{StoneLisberger86,bloedel_current_1998,KitazawaKimuraYin98,YanagiharaUdo94}, 
but it also has been found to correlate with both sensory and motor
events, so that the nature of what is being encoded remains controversial 
\cite{bloedel_current_1998,Anastasio01,Llinas11}.

We have assumed that complex spikes signal an increase in 
error, like the distance between the hand and a target, or the distance between
the hand and its intended point in the trajectory.
This is different from assuming that complex spikes perform a low-frequency encoding of the 
error \cite{SchweighoferDoyaFukaiEtAl04,KitazawaKimuraYin98,KitazawaWolpert05} 
because our onset signal doesn't track the error's magnitude,
it is only related to the positive part of the error's derivative. 
Moreover, this climbing fiber signal does not require high firing rates,
and the magnitude of the error correction could be obtained through several mechanisms
such as cumulative learning through time, graded complex spikes 
\cite{najafi_beyond_2013,yang_purkinje-cell_2014}, or complex 
spike synchrony. 

A noteworthy aspect of our computational simulations when using visual errors 
(models 1 and 4) is
that we included an inferior olivary module that considered a number of units
with subthreshold oscillations. This was done because such a module confers specific
advantages in our architecture.
Our second condition for the integration stop time in the mathematical model
is more likely to be satisfied
when the integration stop time is short. This means that instead of having a single
large correction associated with an error prone area, it may be better to have several
smaller corrections along the trajectory of the arm in state space during episodes
of error increase. Our computational model of the inferior olivary module uses the
subthreshold oscillations of IO cells as a mechanism to generate sequences of complex spikes
during episodes of error increase, instead of having all IO cells firing simultaneously
when there is an increase in error. The increase in error stimulates all IO cells 
targeting a microcomplex, but only those near the peak of their subthreshold oscillation
will respond. As long as the error continues to increase, the IO cells nearing the peak
of their oscillations will tend to activate.

To precisely convey 
the timing of increase onsets and to encourage stability it is important to have 
a wide range of phases in the subthreshold oscillations of inferior olivary 
cells \cite{JacobsonLevYaromEtAl09}, which largely depends on the coupling strength 
of olivary gap junctions \cite{LongDeansPaulEtAl02}. The complex desynchronized 
spiking mode \cite{SchweighoferDoyaKawato99} has a wide range of phases, as
assumed in our simulations. 
We model the subthrehold oscillations of the
IO cells so that the probability to spike for each cell depends on both the
strength of the input signal and the phase of the subthreshold oscillation. 
Larger increases in the error produce stronger input signals to the inferior olivary, 
which are reflected by a larger number of neurons responding; thus, for any short 
time interval, the magnitude of the error increase is reflected by the number of 
inferior olivary cells spiking in synchrony.  The inhibitory feedback from the cerebellar 
nuclear cells, in addition to functioning as a negative feedback system to control 
simple spike discharges \cite{BengtssonHesslow06}, could also help to avoid large 
clusters of synchronized inferior olivary cells, so as to maintain the complex desynchronized 
spiking mode.

\subsection{From DCN activity to behavioral responses}

If the group of Deep Cerebellar Nucleus (DCN) cells in one microcomplex stimulate only one muscle
(or a set of agonists muscles), it is easy to see how in models 2 and 3 the right error signals
for a given microcomplex come from the muscles affected by their DCN cells. In this case 
cerebellar modules can work as 1-dimensional systems,
with an adaptive filter system as the one in \cite{Fujita82} or \cite{chapeau-blondeau_neural_1991} being sufficient to 
perform the corrections. In our simulations, however, the increase in performance in models 2 and 3 was
smaller than that in models 1 and 4, which used visual errors. This, we concluded, was
the cost of using 1-dimensional errors to correct an error that is multidimensional (the distance between
the hand and the target).

On the other hand, models 1 and 4 present a difficulty when considering why the activity of a given 
DCN cell activates the right muscles for a correction. As mentioned before, in models 1 and 4 there is 
only one learning signal (visual error increase), which would be the same for all microcomplexes. 
This implies that all microcomplexes would learn the same response, and entering an error-prone 
region of state space would elicit the responses associated with all DCN cells. Models 1 and 4 specify
what the corrective command is, so conceptually the motor error and redundancy problems are solved, 
but it is worthwhile to think of how this corrective command could become associated with the DCN
activity in the nervous system.
One possible solution to this is to assume that the DCN activity initially only signals the need for a 
correction in a particular system, but it doesn't specify what the correction is. 
There is then an additional step in which the
DCN activity becomes associated with the right correction, as suggested in figure \ref{fig.comp_model4}A.

In the case of model 4, the identity of the right correction is produced at the level of
the spinal cord using the equilibrium lengths from the central controller. In the case of
model 1 the corrections are motor commands, so they will also be available at the spinal cord.
A parsimonious hypothesis is thus that DCN activity becomes associated with corrections
in the spinal cord through temporally asymmetric Hebbian learning. 
This hypothesis thus leads to a model where a group of microcomplexes that produces the same outputs
(because they use the same learning signal), but each microcomplex targets different effectors.
An equivalent model is a single micromplex that targets many different effectors, but its connection
with each effector can learn independently.
In either case the output of a microcomplex is associated with a response only when it happens shortly before 
the region of the spinal cord it innervates becomes active. There are thus two conditions to create a correction:
the context is associated with an error (reflected by the DCN activity), and the effector is associated
with the correction (reflected by the spinal cord activity shortly thereafter).

It has been shown that perceived errors are sufficient to produce adaptation in reaching movements, 
so that executing the corrective motion is not necessary for improving performance 
\cite{KitazawaKohnoUka95,TsengDiedrichsenKrakauerEtAl07}. In its present form, our model 4 may not be sufficient 
to explain these experimental results. On the other hand, as in \cite{Fujita05}, movement execution is not 
necessary to train our first model, as long as shortly after producing an 
error a copy of the subsequent efferent command reaches the cerebellum, even if that command is suppressed.
In the hypothesis of the previous paragraph, however, the motor command reaches the spinal cord,
so depending on the particulars of the temporally asymmetric Hebbian learning suppressing the command
could interfere with learning.

We can mention another hypothesis of how DCN activity (that only signals the
need for a correction, but not the correction) becomes associated with muscle activations. The
hypothesis is that the DCN together with the brainstem and the spinal cord could act like a 
multilayer perceptron that associates the activity of DCN nuclei with muscle activations
that reduce the error. Corrective commands like those of models 1 and 4 permit the creation
of training signals. Although this hypothesis offers great computational flexibility, it is very 
speculative, with many possible variations, so we don't elaborate upon it.

A prediction arising from this discussion is that when using visually generated errors the plasticity at the level of the
brainstem and the spinal cord may be essential for ensuring that the cerebellar corrections achieve their intended effect, at
least during the development period and for the control multiple-jointed limbs. Some models assume that plasticity in
the cerebellum is distributed between the cerebellar cortex and the deep cerebellar nuclei 
\cite{RaymondLisbergerMauk96,garrido_distributed_2013}.
We posit one further memory site outside of the cerebellum, responsible for adjusting the effect of its outputs.
The outputs of cerebellar cortex could both modulate and act as a learning signal for the vestibulum/cerebellar nuclei, while
in turn the output from the cerebellar nuclei could modulate and train the response in the brainstem/spinal cord.

\subsection{Comparison with other models}

A model that is related to the model 1 in this paper was presented by Fujita \cite{Fujita05}. In this model, associative 
learning is used to link motor commands with the subsequent corrections performed by a high-level controller.  
Fujita assumed that if 
a high-level motor center unit made a projection to a microcomplex, then the nuclear cells of that microcomplex and the motor 
center unit would encode the same information. We have no high-level motor center units in our model; instead we have
searched for ways to specifically solve the motor error problem.
Another difference with our model is that the context we associate with a correction may contain afferent information
\cite{GhelarducciItoYagi75,HoltzmanRajapaksaMostofiEtAl06,CasabonaBoscoPerciavalleEtAl10}, and 
allows for the possibility that the same motor command may require different corrections under different circumstances. 

Feedback-error learning \cite{kawato_computational_1992} is a very influential model, 
whose main idea (as illustrated in figure \ref{fig.cb_blocks1}B) is to use the 
output of a feedback controller as the learning signal for an inverse model.
Some of its difficulties were mentioned in the Introduction.
Considering that a feedback controller acts like a linear transformation from
sensory to motor coordinates, the error signal we use in models 1 and 4 (figures 
\ref{fig.comp_model1}, \ref{fig.comp_model4}) is similar to the error signal in motor coordinates 
presented in \cite{kawato_computational_1992} in that it can arise due to error rising in a feedback control 
system, but using sensory coordinates. These sensory coordinates, being part of the control 
loop, are linearly related to the motor coordinates, as the feedback controller is
usually a linear transformation from sensory to motor coordinates. This is consistent with the
fact that both sensory and motor information is present in complex spikes 
\cite{kobayashi_temporal_1998,winkelman_motor_2006}. 

The learning signal in the
recurrent architecture \cite{porrill_recurrent_2007} shown in figure \ref{fig.cb_blocks1}A
is of a different kind, as it is related not directly to the control performance, but
to the prediction errors in a forward model. The forward model in figure \ref{fig.cb_blocks1}A
is predicting the response of the controlled object, whereas the forward model in figure 
\ref{fig.cb_blocks1}C is predicting the response of the central controller.
It should be noted that cerebellar outputs not only target brainstem and spinal cord neurons, 
but also thalamic nuclei that convey their signals to the cerebral cortex. In this sense the 
cerebellar outputs could conceivably be added to both the input and output signals of a
central controller, and how those outputs are used depend on the target structure and
its plasticity mechanisms. It is thus possible that architectures where the cerebellar
output is directed at the input of a brainstem controller ---such as the recurrent
architecture--- could coexist with architectures where the output is added to the motor
commands, such as the architecture in this paper.

Notice that the architecture in figure \ref{fig.cb_blocks1}C , by virtue of being a forward
model that uses 
sensory errors together with a feedback controller is compatible with simple spikes encoding 
sensory errors with both a lead (the future corrections associated with contexts) and a 
lag with the opposite modulation (the sensory error and its associated context is an input 
to Purkinje cells) \cite{popa_predictive_2012,popa_purkinje_2013,popa_cerebellum_2014}. 
It is not clear whether this would be the case in the recurrent architecture of figure 
\ref{fig.cb_blocks1}A, since the input and output of the forward model (motor commands and 
predicted trajectories, respectively) may or may not be associated with sensory errors
(sensory errors would be directly associated with complex spikes).

There are some recent models that specifically address the role of the cerebellum in reaching tasks,
but for the most part they are not concerned with the distal error and redundancy problems.
Some examples are presented next.

In \cite{carrillo_real-time_2008} a relatively realistic cerebellar spiking network was
implemented for real-time control of a 2 DOF robot arm. The arm used open-loop control 
based on calculating a minimum-jerk trajectory that was transformed into a trajectory in
joint angle coordinates, from which crude torque commands were generated. The cerebellum
was capable of reducing the error by providing corrective torques. The redundancy problem
does not arise in this context because their 2 DOF arm moves in a plane, and the elbow
joint does not reach negative angles. Also, the distal error problem is not addressed
since the error of their 4 microzones, each corresponding to a muscle, is provided by the
Inferior Olivary (IO) input based on the difference between the desired and actual trajectories.
Because of the low IO firing rates, a probabilistic encoding has to be used in order to 
communicate this error.

\cite{garrido_distributed_2013} used a cerebellar inverse model to implement
adaptable gain control for a simulated robot arm with 3 DOF performing a smooth pursuit task.
Their model used plasticity at 3 synaptic sites to produce corrective torques at states
that correlated with errors. To represent states Garrido et al. used a granule cell layer
model that generated sequences of binary vectors in discrete time when presented with a
fixed mossy fibre pattern. The activity at Purkinje cells and DCN cells were represented
with scalar values. To solve the distal error and redundancy problems this model is provided
with desired trajectories in intrinsic coordinates. The difference between desired and actual
trajectories is used to calculate errors in each joint by an IO module, which represents
this error as a scalar value. 

In another model \cite{casellato_adaptive_2014} a spiking cerebellar network was used
to implement adaptive control in a real robot. Their model implemented Pavlovian 
conditioning, as well as adaptation in the vestibulo-ocular reflex, and in perturbed
reaching. The redundancy problem and the distal error problem are not addressed, since
their model only controls 1-dimensional responses.

\subsection{Hierarchical control}

An interesting aspect of our architecture comes from its application to hierarchical models such as
Threshold Control Theory (TCT) \cite{FeldmanLevin09,LatashLevinScholzEtAl10}, and Perceptual Control Theory 
(PCT) \cite{powers_feedback:_1973,powers_behavior:_2005}. Briefly, TCT posits that movement control
begins by setting a threshold value for muscle lengths. Muscle contraction happens in response to
the muscle length exceeding this threshold. For a given set of threshold values, interaction with
the environment brings the organism to an equilibrium position; the organism needs to learn the threshold
values that result in desired equilibrium positions. To solve redundancy problems with minimal action,
this paradigm can be extended hierarchically. For example, if there is a neuron that responds montonically
to the aperture of the elbow angle, a controller can set a threshold value for that neuron (the neuron
responds only when the elbow angle goes beyond the threshold). The elbow angle neuron can in turn set the 
threshold lengths of the biceps and triceps brachii muscles so that the its threshold value can in
fact control the elbow angle. At a higher level, there could be neurons that respond to the arm configuration,
and affect the threshold levels for neurons responding to shoulder, elbow, and wrist angles. Each
hierarchical level works as a feedback control system whose set point is specified by the level
above. In this paradigm, known as cascade control, each level isolates the levels above from disturbances
(as long as the lower levels are on a faster timescale than the higher levels),
and redundancies are resolved automatically. PCT shares some of the same ideas as TCT. In PCT the organism
seeks to control its perceptions (instead of TCT's equilibrium positions), and this 
is achieved through a cascade control scheme, going from individual muscles to advanced 
cognitive operations. PCT also proposes a mechanism allowing such a hierarchy of control
systems to arise.  

\begin{figure}
  \centering\includegraphics[height=1.5in]{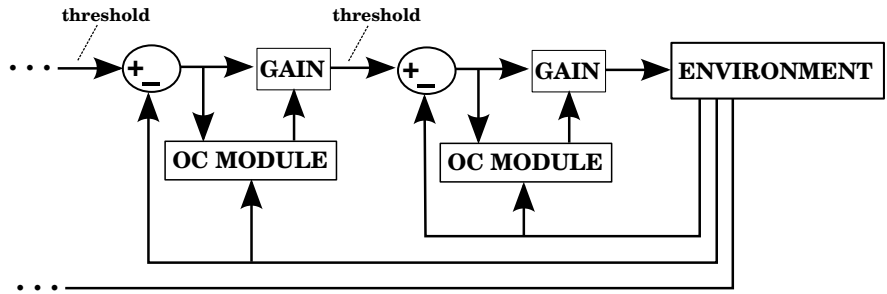}
  \caption{\small Olivo-cerebellar modules used to anticipatively adjust threshold
	  values in a cascade control scheme. The difference between
	  a received threshold value and a value perceived from the environment
	  is transmitted to the olivo-cerebellar module. Increases in this difference
	  cause the olivo-cerebellar module (OC-MODULE) to associate the perceived
	  context at the time of the increase with an anticipative correction. The
	  effect of this correction could be additive, or it could modify a gain on
	  the signal at the GAIN block.
	  Notice that the difference between a threshold value and a perceived value
	  could set the threshold of more than one control loop.
  }
  \label{fig.cb_blocks3}
\end{figure}

Despite their advantages, TCT and PCT rely on feedback control, which can encounter problems in the
presence of time delays and low gains. The cerebellar architecture presented in this paper, based on 
predicting the increase in error, is well suited to improve the performance of TCT or PCT models.
The ideas presented in this paper offer several options to do this. Perhaps the simplest one is to
generate an error signal whenever a threshold value is being exceeded (figure
\ref{fig.cb_blocks3}), similarly to our model 2. The emission of this
error signal can be conditioned on the error increasing on a higher level, similarly to our
model 3. Or similarly to our model 4, the error signal can have its origin on a level high in
the hierarcy, but the corrective signals can be generated at the lower levels using their own
threshold values. This consitutes a hypothesis
of how the cerebellum could function to improve motor and cognitive operations using 
repetitions of the same modular circuit.

\clearpage


\section*{Disclosure/Conflict-of-Interest Statement}
R. C. O'Reilly is CTO at eCortex, Inc., which may derive indirect benefit from the work presented here.

\section*{Acknowledgement}

We thank Tom Kelly and members of the CCN Lab for their input.
Supported by: ARL/GDRS RCTA project under Cooperative Agreement Number W911NF-10-2-0016.


\bibliographystyle{plain}

\end{document}